\documentclass[twocolumn,superscriptaddress,draft,showpacs,showkeys]{revtex4}
\usepackage{bm}
\newcommand*{\ie}[1][]{{\it i.e.}\ }
\newcommand*{\eg}[1][]{{\it e.g.}\,}
\newcommand*{\etc}[1][]{{\it etc.}\,}
\newcommand*{\const}[1][]{{\it const}\,}
\newcommand*{\Jc}[1][]{\mbox{$\mathcal{J}$}}
\newcommand*{\Ld}[1][]{\mbox{$\mathcal{L}$}}

\newcommand*{\R}[1][]{\mbox{$\mathcal{R}$}}
\newcommand*{\half}[1][]{\mbox{\small $\frac{1}{2}$}}
\newcommand*{\dd}[1][]{\mbox{\textrm{d}}}
\newcommand*{\DD}[1][]{\mbox{\textrm{D}}}
\newcommand*{\ptl}[1][]{\partial}

\newcommand{\grad}{\mbox{grad}}
\newcommand{\curl}{\mbox{curl}}

\newcommand*{\al}[1][]{\alpha}
\newcommand*{\ga}[1][]{\gamma}
\newcommand*{\Del}[1][]{\Delta}
\newcommand*{\del}[1][]{\delta}
\newcommand{\eps}{\epsilon}
\newcommand*{\vep}[1][]{\varepsilon}
\newcommand*{\om}[1][]{\omega}
\newcommand*{\Om}[1][]{\Omega}
\newcommand*{\Th}[1][]{\Theta}

\newcommand*{\bom}[1][]{\mbox{\boldmath $\omega$}}
\newcommand*{\bOm}[1][]{\mbox{\boldmath $\Omega$}}
\newcommand*{\bxi}[1][]{\mbox{\boldmath $\xi$}}

\newcommand{\ba}{\mbox{\boldmath $a$}}
\newcommand*{\A}[1][]{\mbox{$\mathcal{A}$}}
\newcommand*{\bA}[1][]{\mbox{\boldmath $A$}}
\newcommand*{\bb}[1][]{\mbox{\boldmath $b$}}
\newcommand*{\bB}[1][]{\mbox{\boldmath $B$}}
\newcommand*{\bE}[1][]{\mbox{\boldmath $E$}}
\newcommand*{\bF}[1][]{\mbox{\boldmath $F$}}
\newcommand*{\G}[1][]{\mbox{$\mathcal{G}$}}
\newcommand*{\bU}[1][]{\mbox{\boldmath $U$}}
\newcommand*{\bn}[1][]{\mbox{\boldmath $n$}}
\newcommand*{\bp}[1][]{\mbox{\boldmath $p$}}
\newcommand*{\bQ}[1][]{\mbox{\boldmath $Q$}}
\newcommand*{\bs}[1][]{\mbox{\boldmath $s$}}

\newcommand*{\bv}[1][]{\mbox{\boldmath $v$}}
\newcommand*{\bV}[1][]{\mbox{\boldmath $V$}}
\newcommand*{\bw}[1][]{\mbox{\boldmath $w$}}
\newcommand*{\bx}[1][]{\mbox{\boldmath $x$}}
\newcommand{\bX}{\mbox{\boldmath $X$}}

\begin{document}
\centerline{\small {\it Physica D} (2007), doi:10.1016/j.physd.2007.09.020}

\title{Variational formulation of the motion of an ideal fluid on the basis 
of gauge principle}
\author{Tsutomu  Kambe}
\email{kambe@ruby.dti.ne.jp}
\homepage{http://www.purple.dti.ne.jp/kambe/}
\affiliation{IDS, Higashi-yama 2-11-3, Meguro-ku, Tokyo, Japan \ 153-0043}
\affiliation{Chern Institute of Mathematics, Nankai University (China), Visiting Professor}

\begin{abstract}
On the basis of gauge principle in the field theory, a new variational formulation 
is presented for flows of an ideal fluid.  The fluid is defined thermodynamically 
by mass density and entropy density, and its flow fields  are characterized by 
symmetries of translation and rotation.   A structure of  rotation symmetry is 
equipped with a Lagrangian $\Lambda_A$ including  vorticity, in addition to Lagrangians
of translation symmetry.  From the action principle, Euler's equation of motion is
derived. In addition, the equations of continuity and  entropy are derived  from the 
variations.  Equations of conserved currents are deduced as the Noether theorem in 
the space of Lagrangian  coordinate $\ba$. It is shown that, with the translation symmetry 
alone, there is freedom in the transformation between the Lagrangian $\ba$-space and  
Eulerian $\bx$-space. The Lagrangian $\Lambda_A$ provides non-trivial topology of vorticity 
field and yields a source term of the helicity.  The vorticity equation is derived as 
an equation of the gauge field. Present formulation provides a basis on which the 
transformation between the  $\ba$ space and the  $\bx$ space is determined uniquely.  
\end{abstract}

\pacs{47.10.-g, 03.50.-z, 47.15.ki, 47.90.+a}
\keywords{Variational formulation, Gauge principle, Euler's equation, Helicity,
Chern-Simons term}

\maketitle

\section{Introduction \label{S1-Int}}

In the historical paper '{\it General laws of the motion of fluids}' \cite{Euler1755}, 
Leonhard Euler verified that his equation of motion can describe rotational flows. 
The same theme is investigated in this paper under modern view. {\it Fluid mechanics} is 
understood to be a field theory in Newtonian mechanics that has Galilean symmetry.  
It is covariant under transformations of the Galilei group.  The gauge principle 
(\cite{We95}, \cite{Fr97}, \cite{AH82}) requires a physical 
system under investigation to have a {\it symmetry}, \ie a gauge invariance  with respect 
to a certain group of transformations.  Following this principle, the gauge symmetry of 
flow fields  is studied in \cite{Ka07a} and \cite{Ka07b} with respect to both translational 
and rotational transformations.  The formulation started from a 
Galilei-invariant Lagrangian of a system of {\it point masses} which is known to have 
{\it global} gauge symmetries with respect to both translation and rotation \cite{LL76}.  
It was then extended to flows of a fluid,  a continuous material characterized 
with mass density and entropy density. In addition to the global symmetry, {\it local} 
gauge invariance of a Lagrangian is required for such a continuous field. Symmetries imply 
conservation laws. Equations of conserved currents are deduced as the Noether theorem.

Thus, the convective derivative of fluid mechanics, \ie the Lagrange derivative, is 
identified as the {\it covariant derivative}, which is a building block in the framework 
of gauge theory.  Based on this, appropriate Lagrangians are defined  for 
motion of an ideal fluid.  Euler's equation of motion is derived from the 
action principle.   In most traditional formulations, the continuity equation and  
entropy equation are given as constraints for the variations, while in this new 
formulation those equations were derived from the action principle.
In the previous study (\cite{Ka03a}, \cite{Ka03b}) of rotational symmetry of the velocity 
field $\bv(\bx)$, it is found that the vorticity $\bom=\nabla\times \bv$ is  
the gauge field associated with the rotational symmetry of velocity.  

A new structure of the rotational symmetry was given in \cite{Ka07b} by the
following Lagrangian:
\[	\Lambda_A = - \int_{M} \langle \Ld_W \bA,\, \bom \,\rangle \,\dd^3\bx ,  \]  
where $\bA$ is a vector potential and $\Ld_W \bA = \ptl_t \bA + v^k \ptl_k \bA
+ A_k \nabla v^k$.  This is derived 
from a representation characteristic of a {\it topological} term known in the gauge 
theory.  This yields non-vanishing rotational component of the velocity field, and 
provides a source term of helicity. This is closely related to the Chern-Simons
term, describing non-trivial topology of vorticity field, 
\ie mutual linking of vorticity lines.  The vorticity equation is derived as an equation 
for the gauge field. 

With regard to the variational formulation of fluid flows, the papers  \cite{Ec60} 
and \cite{He55} are among the earliest to have influenced current formulations.  Their 
variations are carried out in two ways: \ie a Lagrangian approach and an Eulerian approach.
  In both approaches, the equation of continuity and the condition of isentropy are added 
as constraint conditions on the variations by means of Lagrange multipliers.  The 
Lagrangian approach is also taken by \cite{So76}. In this relativistic formulation those 
equation  are derived from the equations of current conservation.  Several action 
principles to describe relativistic fluid dynamics have appeared in the past 
(see \cite[\S4.2]{So76} for some list of them).

In the  Lagrangian approach,  the Euler-Lagrange equation results in an equation equivalent 
to Euler's equation of motion in which the acceleration term is represented as the second 
time derivative of position coordinates of the Lagrangian representation.  In this 
formulation, however, there is a certain degree of freedom in the relation between the 
Lagrangian particle coordinates and Eulerian space coordinates. Namely, the relation between 
them is determined only up to an unknown rotation.  In the second approach referred to as 
the {\it Eulerian} description, the action principle of an ideal fluid results in potential 
flows with vanishing helicity, if the fluid is {\it homentropic} \cite{Ka07a}. 
However, as noted in the beginning, it should be possible to have rotational flows 
even in such a homentropic fluid.  Gauge theory for fluid flows  provides  a crucial key 
to resolve these issues.  It was also shown in \cite{Ka07a} that a general solution in the 
translational symmetry alone is equivalent to the classical Clebsch solution \cite{Cl1859}. 
A new formulation on the basis of the  Clebsch parametrization is carried out in \cite{Ja02}
and \cite{Ja04} aiming at its extension to supersymmetric and non-Abelian fluid mechanics.

It is interesting to note  the gauge invariances known in the theory of electromagnetism 
and fluid flows.  There is an invariance under a gauge transformation of electromagnetic 
potentials consisting of a scalar potential $\phi$ and a vector potential $\bA$.  
An analogous invariance is pointed out in \cite{Ka07a} for a gauge transformation of 
a velocity potential $\phi$  of irrotational flows of an ideal fluid, where the velocity 
is represented as $\bv= \nabla \phi$.  It is shown in \cite{Ec60} (cited in \cite{Ka07b}) 
that gauge invariance is not restricted to the potential flows, but also there is known 
an invariance in the rotational flow of  {\it Clebsch representation}.

\section{Equations in  $\ba$-space \label{2-L-a} }

\subsection{Lagrangian \label{2A}}

Let us consider a variational formulation with a Lagrangian  represented with the 
particle coordinate $\ba=(a^1,a^2,a^3)=(a,b,c)$ (\ie {\it Lagrangian} coordinates).  
Independent variables are denoted with $a^\mu$ where $\mu$ or greek letter suffix
take $=0,1,2,3$ with $a^0$  the time variable written also as $\tau$ ($=t$):  $a^\mu 
=(\tau, a^1,a^2,a^3)$.  Corresponding physical space coordinate $\bx=(x,y,z)$ ({\it 
Eulerian} coordinates) are written also as $x^\mu =(t, x^1,x^2,x^3)$. The letter $\tau$ 
is used (instead of $t$) in combination with the particle coordinates $a^k$.   Physical
space position of a particle $\ba$ is expressed by $X^k(a^\mu)=X^k(\tau, \ba)$, or 
$X^k=(X,Y,Z)$.  Its velocity is given by $v^k = \ptl_\tau X^k$, also written as $X_\tau^k$. 

The Lagrangian coordinates $(a,b,c)$ are defined such that an infinitsimal 
three-element $\dd^3 \ba =\dd a \,\dd b \,\dd c$  denotes a mass element $\dd m$ of an 
infinitesimal volume  $\dd^3\bx=\dd x \,\dd y \,\dd z$ of the $\bx$-space.  The mass 
elelment $\dd m$ should be invariant during the motion: 
\begin{equation}
\ptl_\tau (\dd m) \equiv \ptl_\tau (\dd^3 \ba) = 0  \,. \label{dt-dm}
\end{equation}
The mass-density $\rho$ is defined by the equation $\dd^3 \ba= \rho\,\dd^3\bx$. 
With using a Jacobian determinant $J$ of the transformation $X^k=X^k(a^l)$ from 
$\ba$-space to $\bX$-space ($k,\,l=1,2,3$),  we have 
\begin{equation}
\rho = \frac{1}{J}, \hskip5mm J = \frac{\ptl(X^1,X^2,X^3)}{\ptl(a^1,a^2,a^3)}
	= \frac{\ptl(X, Y, Z)}{\ptl(a, b, c)}\,. \label{Jacob}
\end{equation}
In an ideal fluid, there is no dissipation of kinetic energy into heat, by 
definition.  According to thermodynamics for the  entropy $s$ (per unit mass) and 
temperature $T$,  we have $T \del s=0$ if there is no heat production. Namely 
the entropy $s$ does not depend on $\tau$.  Then, the change of internal energy 
$\eps$ (per unit mass) is related to the density change $\del \rho$ alone by
\begin{equation}
\del \eps = (\del \eps)_s = \frac{p}{\rho^2}  \,\del \rho, \hskip4mm 
\big( \frac{\ptl \eps}{\ptl \rho}\big)_s = \frac{p}{\rho^2}, 
\hskip4mm \del h = \frac{1}{\rho} \del p, 		\label{del-e}
\end{equation}
where  $p$ is the fluid pressure, and $h=\eps+p/\rho$  the enthalpy, 
and $(\,\cdot\,)_s$ denotes  $s$ being fixed.  
However, the entropy $s$ may not be uniform and may depend on $\ba$ 
by initial condition.  Hence, $s=s(\ba)$, or equivqalently,
\begin{equation}
\ptl_\tau \,s  = 0  \,. \label{dt-s}
\end{equation}
Total Lagrangian is defined by
\begin{equation}
\Lambda_{\rm T} = \int_{M_a} \half\, X_\tau^k\,X_\tau^k\, \dd^3\ba 
	- \int_{M_a} \eps(\rho,s)\, \dd^3\ba \,, 	\label{LT-a}
\end{equation}
\cite{Ka07b}, where $M_a$ is a space of fluid under investigation, and 
$X_\tau^k =X_0^k = v^k$ is the velocity. The internal energy $\eps(\rho,s)$ 
of the second term depends on $\rho$ (which in turn depends on 
$X^k_l=\ptl X^k/\ptl a^l$ by (\ref{Jacob})) and the entropy $s(\ba)$. 

An action  $I$ is defined by the integral: $I=\int \Lambda_{\rm T} \dd\tau$:
\begin{eqnarray}
I & = &  \int\, L(X_\mu^k) \ \dd^4 a,  \hskip8mm \dd^4 a= \dd \tau\,\dd^3\ba,
		   \label{I-LT-a} \\
 &  &   L(X_\mu^k) = \half\,X_0^k\,X_0^k - \eps(X^k_l,a^k).    \label{Lag1}
\end{eqnarray}

\subsection{Noether's theorem \label{2B}}

Euler-Lagrange equation associated with the Lagrangian (\ref{Lag1}) is given by
\begin{equation}
\frac{\ptl}{\ptl a^\mu} \Big( \frac{\ptl L}{\ptl X^k_\mu} \Big) - \frac{\ptl L}{\ptl X^k} 
= \ptl_\mu \Big(\frac{\ptl L}{\ptl X^k_\mu} \Big) -\frac{\ptl L}{\ptl X^k} =0. \label{EL-X}
\end{equation} 
Energy-momentum tensor $T_\mu^\nu$ is defined by 
\begin{equation} 
T_\mu^\nu \equiv X_\mu^k\, \Big( \frac{\ptl L}{\ptl X_\nu^k} \Big) 
		- L\,\del_\mu^\nu \,,  \label{EM-tensor}
\end{equation} 	
\cite{Ec60}, where $k=1,2,3$. As long as (\ref{EL-X}) is satisfied together with an 
assumption of $\tau$-independence of $L$ (\ie $\ptl_\tau L=0$), it can be verified 
\cite{Ka07b} that we have a conservation equation $\ptl_\nu T_\mu^\nu=0$ (where 
$\ptl_\mu=\ptl/\ptl a^\mu$).  This is the Noether theorem (\cite{No18}, \cite{We95}).

For $\mu \ne 0$ ($x^\mu=\al$), the conservation law $\ptl_\nu T_\mu^\nu=0$ 
reduces to the momentum equaitons: 
\begin{equation}
\ptl_\tau V_{\al} + \ptl_{\al} \,F = 0  \hskip4mm 
(V_{\al} \equiv X_{\al} X_\tau + Y_{\al} Y_\tau + Z_{\al} Z_\tau), \label{EqM-a-1}
\end{equation}     
\cite{Ec60}, where $F=- \half\,v^2 + h$.  Two other equations are obtained with $\al$ 
replaced by cyclic permutaion of $(a,b,c)$. Integrating this with respect to $\tau$ 
between 0 and $t$, we find the Weber's transformation \cite[Art.15]{La32}:
\begin{eqnarray} 
V_{\al}(\tau) & \equiv & X_{\al} X_\tau + Y_{\al} Y_\tau + Z_{\al} Z_\tau  
		= V_{\al}(0) - \ptl_{\al} \chi, \hskip3mm  \label{Web-Tr}	\\
 && \chi = \int_0^t \, F\,\dd\tau = \int_0^t (- \half\,v^2 + h) \dd\tau. \nonumber
\end{eqnarray} 
The $V_{\al}$ of (\ref{EqM-a-1}) is  a transformed velocity in the 
$\ba$-space (Sec.\ref{5A}).  Its time evolution is given by   (\ref{Web-Tr}) for 
a given initial values of $V_{\al}(0,\ba)$ and $h(0,\ba)$ at $\ba=\bx$.  

With $\mu=0$, we have the energy equation:
\begin{eqnarray}
\ptl_\tau H & + & \ptl_a \Big[ p \frac{\ptl (X, Y, Z)}{\ptl (\tau, b, c)} \Big]
   + \ptl_b \Big[ p \frac{\ptl (X, Y, Z)}{\ptl (a, \tau, c)} \Big] \nonumber \\
 & + &  \ptl_c \Big[ p \frac{\ptl (X, Y, Z)}{\ptl (a, b, \tau)} \Big] = 0.  \label{CEq-0}
\end{eqnarray}
where $H= \half\,v^2 + \eps$.  The equation (\ref{EqM-a-1}) reduces to the equation 
for the acceleration $\A_{\al}(\tau,\ba)$:
\begin{equation}
\A_{\al} \equiv X_{\al} X_{\tau\tau} + Y_{\al} Y_{\tau\tau} + Z_{\al} Z_{\tau\tau} 
	= - \frac{1}{\rho}\,\ptl_{\al} p ,  \label{EqM-a-2}
\end{equation}
which is known as the Lagrangian form of equation of motion \cite[Art.13]{La32}.
This can be transformed to 
\begin{equation}
X_{\tau\tau} = - \frac{1}{\rho}\,\ptl_x\,p , \hskip10mm \ptl_x p = 
	\frac{\ptl \al}{\ptl x} \,\frac{\ptl p}{\ptl \al}      \label{EqM-x-1}
\end{equation}
\cite{Ka07b}.  Since $X_{\tau\tau}$ is the $x$-accceleration of the particle $\ba$, 
this is the form equivalent to the $x$-component of Euler's equation of motion 
(\ref{EEq}). The $y$ and $z$ components can be obtained analogously.

\subsection{Arbitrariness in the transformation \label{2C}}

There is an arbitrariness in the transformation from the $\ba$-space to the $\bx$-space 
with respect to the equation (\ref{EqM-a-2}). Its middle-side expression is a form of 
scalar product of two vectors in the $\bx$-space:  the particle accelerataion  
$(X_{\tau\tau}, Y_{\tau\tau}, Z_{\tau\tau})$ and the direction vector 
$(X_{\al}, Y_{\al}, Z_{\al})$  of the $\al$-axis in the $\ba$-space. 

Putting it in a different way, the equation (\ref{EqM-a-2}) is invariant with 
respect to orthogonal rotational transformations of a displacement vector $\Del \bX
=(\Del X, \Del Y, \Del Z)$ of a particle in the $\bx$-space.  In fact, suppose 
that a vector $\Del \bX$ satisfies the equation (\ref{EqM-a-2}).  Then, another 
vector $\overline{\Del \bX}=R\ \Del \bX$ obtained by an orthogonal transformation $R$ 
satisfies the same equation, since  any orthogonal matrix satisfies $RR^T=I$ (unit
mtrix) where $R^T$ denotes the transposed matrix of $R$.  So that the vector $\Del\bX$ 
is not uniquely determined.  The same freedom can be said to the velocity 
$V_{\al}(\tau,\ba)$ of (\ref{Web-Tr}) as well. 

These imply that a certain machinery must be equipped in order to fix this arbitrariness
within the framework of rotational symmetry.  This will be  considered later.  Note 
that the density $\rho$ is not changed by the orthogonal transformation.

\section{Equations in  $\bx$-space \label{3-Eqs-x} }

\subsection{Action in Eulerian representation \label{3A}}
 
Eulerian description is represented by the independent variables $(t,x,y,z)$.  Local 
gauge symmetries of fluid flows are investigated in detail in \cite{Ka07a}, \cite{Ka07b}. 
The time derivative $\ptl_\tau$ is  equivalent to the convective derivative $\DD_t$:
\begin{equation}
\ptl_\tau = \DD_t, \hskip3mm \DD_t \equiv \ptl_t + u\ptl_x + v\ptl_y +w\ptl_z 
= \ptl_t + \bv\cdot\nabla \,. \label{dt-Dt}
\end{equation}
The operator $\DD_t$ is verified to be gauge-invariant.  The velocity field $\bv(\bx,t)$ 
is defined by the particle velocity:
\begin{equation}
\bv(\bx,t) = \ptl_\tau \bX = \DD_t \bx.		   \label{v-Dtx} 
\end{equation}
The acceleration field $\A(\bx,t)$ is also defined by 
\begin{equation}
\A(\bx,t) = \ptl_\tau^{\ 2} \bX = \DD_t \bv = (\ptl_t  + v^k \ptl_k) \bv.   \label{a-Dt2x} 
\end{equation}
As noted previously,  the mass  $\dd^3 \ba(\ba)$ and the entropy  $s=s(\ba)$ satisfy 
(\ref{dt-dm}) and (\ref{dt-s}). In view of these properties, we can define the 
following two Lagrangians:
\begin{equation}
L_\phi = - \int_{M}\,\ptl_\tau \phi \,\dd^3\ba, \qquad
L_\psi = - \int_{M}\,s\,\ptl_\tau \psi \, \dd^3\ba,      \label{Lph-ps-a}
\end{equation}
where $\phi(\ba,\tau)$ and  $\psi(\ba,\tau)$ are scalar fields associated with
mass and entropy, respectively. By adding $L_\phi$ and $L_\psi$ to 
$\Lambda_{\rm T}$ of (\ref{LT-a}), the total Lagrangian is given by 
\begin{equation}
\Lambda_{\rm T}^{\ *} = \Lambda_{\rm T} - \int \ptl_\tau\phi\ \dd^3\ba 
	- \int s\, \ptl_\tau\psi\ \dd^3\ba. \label{LT-2}
\end{equation} 
The action is defined by  $I=\int_{\tau_1}^{\tau_2} \Lambda_{\rm T}^{\ *} \dd\tau$, 
where  the integral $I_\phi =\int \dd\tau \int \ptl_\tau\phi \,\dd^3\ba$ 
can be integrated with respect to $\tau$ and expressed as $\int [\phi] \dd^3\ba$, where 
$ [\phi]= \phi|_{\tau_2} - \phi|_{\tau_1}$ is the difference of $\phi$ at the end times 
$\tau_2$ and $\tau_1$ and hence independent of $\tau \in (\tau_1,\tau_2)$.  Likewise, the 
last integral can be expressed as $I_\psi=\int [\psi] s\, \dd^3\ba$, because  $s$ is  
independent of $\tau$. This means that the gauge potentials $\phi$ and $\psi$ do not 
appear in the equation of motion obtained through variations of the action 
$I$  for $\tau \in (\tau_1,\tau_2)$. 

However, it becomes  soon clear that these are nontrivial in the expressions of the 
$\bx$-space,  because they are rewritten as $L_\phi =- \int_{M}\rho\,\DD_t\phi\,
\dd^3\bx$, and $L_\psi =-\int_{M} \rho s\,\DD_t\psi \,\dd^3\bx$  by using the 
relations $\dd^3 \ba= \rho\,\dd^3\bx$ and $\ptl_\tau=\DD_t$.

In the $\bx$-space,  the total Lagrangian can be written  as $\Lambda_{\rm T}^{\ *} = 
\int_{M} \Ld(\bv, \rho, s, \phi,\psi)\ \dd^3\bx$, where 
\begin{equation}
\Ld   \equiv   \half \rho\, v^k v^k - \rho \eps(\rho,s)  
	- \rho \DD_t\phi - \rho s \DD_t\psi  	\label{LT-x-1}
\end{equation}
\footnote{Roman Jackiw \cite{Ja02} arrived at the same form, but with a different approach 
using Lagrannge multipliers for constraint conditions, in order to extend it
to the relativistic case, {\it e.g.} Eq.(2.56) of \cite{Ja02} 
and (1.2.68) of \cite{Ja04}.}.
This is proposed as a {\it possible} form of Lagrangian in the $\bx$-space (but an 
additional term will be added later).  The action is defined by  $I = \int  
\Ld (\bv, \rho, s, \phi,\psi) \ \dd^4 x $,  where $\dd^4 x= \dd t\,\dd^3\bx$.
However, the action principle  results  in the potential flow represented by 
$\bv=\grad(\phi+ s_0 \psi)$ when the fluid has a uniform entropy $s_0$ (see \cite{Ka07a}).

\subsection{Outcomes of variations \label{4A}}

We require invariance of the action $I$ with respect to variantions. First, 
consider  the following infinitesimal  transformation: $ \bx'(\bx,t)= \bx 
+ \bxi(\bx, t)$. The volume element $\dd^3\bx$ is changed to $\dd^3\bx'=(1+ \ptl_k\xi^k)
\dd^3\bx$, up to the first order terms. Hence the variation of volume  is given by 
$\Del(\dd^3\bx) = \ptl_k \xi^k\, \dd^3\bx$, while the variations of density, velocity  and 
entropy are $\Del \rho =- \rho\ \ptl_k\xi^k$,  $\Del \bv = \DD_t \bxi$, and 
$\Del s = 0$. 	Under these  together with (\ref{dt-dm}) and (\ref{dt-s}) 
(with keeping $\phi$ and $\psi$ fixed), the variation of $I$ is given by
\[  
\Del I = \int \dd^4x\, \Big[ \, \frac{\ptl L}{\ptl \bv} \Del \bv + \frac{\ptl L}{\ptl \rho} 
  \, \Del \rho + \frac{\ptl L}{\ptl s} \,\Del s + L\,\ptl_k \xi^k \,\Big] . \label{del-Ix}  
\] 
This  is required to vanish for arbitrary variation of $\xi^k$, which results in 
the  Euler-Lagrange equation:
\begin{equation}
\frac{\ptl}{\ptl t} \big( \frac{\ptl L}{\ptl v^k} \big) + \frac{\ptl}{\ptl x^l} 
	\big( v^l\, \frac{\ptl L}{\ptl v^k} \big) + \frac{\ptl}{\ptl x^k} \big( L 
	- \rho \, \frac{\ptl L}{\ptl \rho}  \big) = 0 .		\label{ELeq-Fluid1} 
\end{equation}
Similarly, invariance of $I$ with respect to arbitrary variations of $\phi$ and 
$\psi$ (denoted by $\Del \phi$ and $\Del \psi$) leads to 
\begin{eqnarray}
  \Del \phi   &:&  \ \ptl_t \rho + \nabla\cdot(\rho \bv) =0  \hskip2mm	
\mbox{({\it continuity equation})},  \hskip4mm   \label{dphi} \\ 
  \Del \psi   &:&  \ \ptl_t(\rho s)+ \nabla\cdot(\rho s \bv) =0\,.	\label{dpsi1}
\end{eqnarray}

\subsection{Noether's theorem in Eulerian representation \label{4C}}

Associated with (\ref{ELeq-Fluid1}),  one can define the momentum density $m_k$ and 
momentum-flux tensor $M_k^l$  by 
\begin{equation} 
m_k = \frac{\ptl L}{\ptl v^k}, \hskip5mm M_k^l = v^l\,\frac{\ptl L}{\ptl v^k} 
   + \big( L - \rho\,\frac{\ptl L}{\ptl \rho}\big) \,\del_k^l \,. \label{m-Mflux-1}
\end{equation} 	
From (\ref{Lag1}), we obtain $m_k = \rho v_k$ and $M_k^l = \rho v_kv^l + p\,\del_k^l$, 
where $v_k=v^k$ in the present Eucledian space.  The equation (\ref{ELeq-Fluid1}) can 
be written in the form of momentum conservation, $\ptl_t \big(\rho v^k \big) 
+ \ptl_l \big(\rho v^l v^k \big) + \ptl_k p= 0$  ($\ptl_k=\ptl/\ptl x^k$).  Using  
(\ref{dphi}),  this equation can be reduced to  the following Euler's equation of motion:
\begin{equation}
\ptl_t v^k + (v^l \ptl_l) v^k = - \frac{1}{\rho}\,\ptl_k\,p \quad (= -\ptl_k\,h). \label{EEq}
\end{equation}  
The equation (\ref{EqM-x-1}) is equivalent to this equation.

The energy equation (\ref{CEq-0})  can be transformed to the following equation of 
energy conservation:
\[  \ptl_t \left[ \rho (\half \,v^2 + \eps) \right] 
	+ \ptl_k \left[ \rho v^k\,(\half \,v^2 +h) \right]=0.   \]

\section{Rotation symmetry \label{S4}}

A topological structure of vorticity field is now considered with respect to the 
rotational symmetry.  Related gauge group is the rotation group $SO(3)$.  An 
infinitesimal rotation is described by the Lie algebra  ${\bf so}(3)$ of three 
dimensions, which is non-Abelian. 

From the study  of the rotational gauge transformation \cite{Ka07b}, it is found that 
the covariant derivative $\nabla_t$, velocity $\bv$ and accerelation $\A$ are represented as 
\begin{eqnarray}
\nabla_t  & = & \ptl_t  + (\bv \cdot \nabla),		\label{R-CD}  \\
\bv = \nabla_t \bx & = & (\ptl_t  + (\bv \cdot \nabla))\bx,  \label{R-vel}  \\
\A = \nabla_t \bv & = & \ptl_t \bv + (\bv \cdot \nabla)\,\bv  \label{R-acc1}  \\
\nabla_t \bv  & = &  \ptl_t \bv + \grad(\half\, v^2) + \bom \times \bv.  \label{R-acc2}
\end{eqnarray}
It is verified  that the last expression of $\nabla_t \bv = \ptl_t \bv + 
\nabla(\half v^2) + \bom \times \bv $ not only satisfies the rotational gauge-invariance,
but also expresses that $\bom$ is the gauge field of the rotational symmetry. 
In addition, it satisfies the covariance requirement with respect to  Galilean 
transformation from one reference frame ($t,\,\bx,\,\bv$) to another ($t_*,\,\bx_*,
\,\bv_*$) moving with  a uniform relative velocity $\bU$, where $t_*=t$, $\bx_* =
\bx - \bU t$ and $\bv_*=\bv - \bU$. Namely, we have the covariance 
$\nabla_t \bv = (\nabla_t \bv)_*$. 

\section{Lagrangian associated with rotation symmetry \label{S7}}

Associated with  the rotation symmetry, an additional Lagrangian is to be defined 
according to the  gauge principle.   It is important to observe from  Sec.\ref{3A} 
that, in the Lagrangian (\ref{LT-2}), the integrands of the last two 
integrals are of the form $\ptl_\tau(\,\cdot\,)$.  The action is defined by 
$I=\int\int [\Lambda_{\rm T}+ \ptl_\tau(\,\cdot\,)]\,\dd\tau \dd^3\ba$.
This property is regarded as the simplest  representation of topology in the gauge theory 
(\cite{Ch79} $\sim$ \cite{DJT82}, \cite{We95}). In the context of rotational flows, it is 
known that the helicity (or Hopf invariant, \cite{AK98}) describes non-trivial topology 
of vorticity field, \ie mutual linking of vorticity lines.  This is closely related with 
the Chern-Simons term (without third-order term) in the gauge theory. This term lives in one 
dimension lower than the original four space-time ($x^\mu$) of the action $I$ because a 
topological term in the action is expressed in a form of total divergence ($\ptl_\mu F^\mu$) 
and characterizes topologically non-trivial structures of the gauge field.

However, we learn here from the formulation of Sec.\ref{3A} and  look for a 
$\tau$-independent field  directly.

\subsection{Lagrangian $\Lambda_A$ and helicity   \label{5A}}

The $\tau$-independent field   can be found immediately from Eq.~(\ref{EqM-a-1}).  
Taking the curl of this equation with respect to the coordinates $(a,b,c)$, we  obtain 
\begin{equation}
\nabla_a \times \ptl_\tau \bV_a = \ptl_\tau (\nabla_a \times \bV_a)= 0 , \label{dtOm-a}
\end{equation}
where $\nabla_a=(\ptl_a, \ptl_b, \ptl_c)$.  Hence, one may write as 
$\nabla_a\times \bV_a = \bOm_a(\ba)$ \cite{Ec60}.  

The vector $\bV_a$ is  a transformed form of the velocity $\bv=(X_\tau,Y_\tau,Z_\tau)
=(u,v,w)$  into the $\ba$-space.  This is seen on the basis of a 1-form $V^1$ defined by
\begin{eqnarray}
V^1      & = & V_a \,\dd a + V_b \,\dd b + V_c \,\dd c    \label{V1aa} \\
        & = & u\,\dd x + v \, \dd y + w \,\dd z. \hskip5mm \label{V1xx} 
\end{eqnarray}  
where $V_a= u x_a + v y_a + w z_a$, $x_a=\ptl X/\ptl a$, $u=X_\tau$, \etc.
Its differential $\dd V^1$ gives a two-form $\Om^2 = \dd V^1$:
\begin{eqnarray} 
\hspace*{-5mm} \Om^2     & = & \Om_a  \dd b \wedge \dd c + \Om_b \dd c 
     \wedge \dd a + \Om_c \dd a \wedge \dd b    \nonumber  \\
   & = & \om_x \dd y \wedge \dd z + \om_y  \dd z \wedge \dd x
		+ \om_z \dd x \wedge \dd y,  \label{Om2sS}	
\end{eqnarray}
where $(\Om_a, \Om_b,\Om_c)=\bOm_a$, and  $\nabla\times \bv
= (\om_x, \om_y, \om_z)=\bom$ is the vorticity.  Thus, it is seen that $\bOm_a$
is the vorticity transformed to the $\ba$-space. The equation (\ref{dtOm-a}) is 
transformed into the $\tau$-derivative  of the 2-form $\Om^2$, $\Ld_{\ptl_\tau} \Om^2=0$
(understood as the  Lie derivative).

Next, let us introduce a gauge-potential vector $\bA_a=(\overline{A}_a,\overline{A}_b,\overline{A}_c)$ in the $\ba$-space,  
and define  its 1-form $A^1$ by $A^1  = \overline{A}_a\,\dd a + \overline{A}_b\, \dd b + \overline{A}_c\,\dd c  
= \overline{A}_x\,\dd x + \overline{A}_y\, \dd y + \overline{A}_z\,\dd z$.   Thus, it is  proposed that a {\it possible} 
type of Lagrangian is 
\[ \Lambda_A =  - \int_{M} \langle \ptl_\tau \bA_a, \bOm_a \rangle \,\dd^3\ba   
	=   \int_{M} \langle \bA,\, E_W[\bom] \,\rangle \,\dd^3\bx,  \] 
where $E_W[\bom] \equiv \ptl_t \bom  + (\bv\cdot\nabla)\bom -(\bom\cdot\nabla)\bv 
+ (\nabla\cdot \bv) \bom$.  

New results were deduced from this Lagrangian in \cite{Ka07b}: (i) the velocity $\bv$ 
includes a new rotational term, (ii) the vorticity equation is derived from the variation 
of $\bA$: 
\[  E_W[\bom] = \ptl_t \bom  + (\bv\cdot\nabla)\bom -(\bom\cdot\nabla)\bv 
	+ (\nabla\cdot \bv) \bom =0, 	\]
and (iii)  we have non-vanishing helicity $H$, where  
\[  H =  \int_V \ \bom \cdot \bv \ \dd^3\bx 
	= \int_V   \bom \cdot \frac{E_W[\curl \bA]}{\rho} \, \dd^3\bx,  \]

\subsection{Uniqueness of transformation   \label{9B}}

Transformation from the  Lagrangian $\ba$ space to Eulerian $\bx(\ba)$ space is 
determined locally by nine components of the matrix $\ptl x^k/\ptl a^l$. However, in the 
previous solution considered in Sec.\ref{2C},  we had three relations (\ref{Web-Tr}) 
between $\bv=(X_\tau, Y_\tau, Z_\tau)$ and $(V_a, V_b,V_c)$, and another three relations 
(\ref{EqM-a-2}) between $\A=(X_{\tau\tau}, Y_{\tau\tau}, Z_{\tau\tau})$ and $(\A_a, \A_b,\A_c)$. 
Remaining three conditions are given by the equation (\ref{Om2sS}) connecting 
 $\bom=(\om_x, \om_y, \om_z)$ and $\bOm_a(\ba)=(\Om_a, \Om_b,\Om_c)$.  
For example, $\Om_a$ is determined by 
\begin{eqnarray}  
\Om_a & = & \om_x\,(\ptl_b y\,\ptl_c z - \ptl_c y\,\ptl_b z) 
		+ \om_y\,(\ptl_b z\,\ptl_c x - \ptl_c z\,\ptl_b x)  \nonumber  \\
  && \hspace*{-3mm} + \ \om_z\,(\ptl_b x\,\ptl_c y - \ptl_c x\,\ptl_b y). \label{omx-Oma}
\end{eqnarray}

There are three vectors (velocity, acceleration and vorticity) determined by 
evolution equations subject to initial conditions in each space of $\bx$ and $\ba$ 
coordinates.  Transformation relations of the three vectors suffice to determine
the nine matrix elemets $\ptl x^k/\ptl a^l$ locally. Thus, the transformation between the  
Lagrangian $\ba$ space and Eulerian $\bx(\ba)$ space is determined uniquely. \cite{Ka07b}

\section{Summary and discussion}

Following the scenario of the gauge principle of field theory, it is found that 
the variational principle of fluid motions can be reformulated successfully in terms 
of covariant derivative and  Lagrangians defined appropriately.  
The present variational formulation is self-consistent and comprehensively 
describes flows of an ideal fluid.

In the improved formulation taking account of the rotational symmetry with additional 
equations of (\ref{Om2sS}), the transformation relations of the three vectors 
(velocity, acceleration and vorticity) suffice to determine the nine matrix elements 
$\ptl x^k/\ptl a^l$ locally. Thus, the transformation between the  Lagrangian $\ba$ 
space and Eulerian $\bx(\ba)$ space is determined uniquely.


\bibliography{EE250sample}

\end{document}